# Polymer Nanofibers and Nanotubes: Charge Transport and Device Applications

**By Andrey N. Aleshin**[*]

*A. F. Ioffe Physical-Technical Institute, Russian Academy of Sciences, St. Petersburg 194021, Russia and School of Physics & Nano Systems Institute - National Core Research Center, Seoul National University, Seoul 151-747, Korea*

**Abstract**

A critical analysis of recent advances in synthesis and electrical characterization of nanofibers and nanotubes made of different conjugated polymers is presented. The applicability of various theoretical models is considered in order to explain results on transport in conducting polymer nanofibers and nanotubes. The relationship between these results and the one-dimensional (1D) nature of the conjugated polymers is discussed in light of theories for tunneling in 1D conductors (e.g. Luttinger liquid, Wigner crystal). The prospects for nanoelectronic applications of polymer fibers and tubes as wires, nanoscale field-effect transistors (nanoFETs), and in other applications are analyzed.

[*] Correspondent author. E-mail: aleshin@transport.ioffe.ru

**Table of Contents**

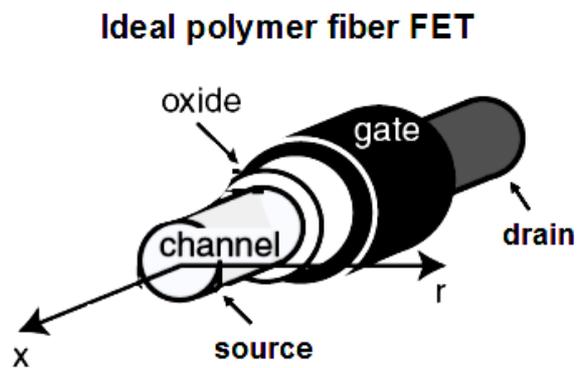

Figure for Table of contents. Ideal polymer fiber FET

## 1. Introduction

Recent remarkable developments in scaling semiconductor nanotechnology have almost reached the limits of physical miniaturization for electronic devices made of conventional semiconductors. The ultimate channel lengths in Si complementary metal-oxide semiconductor (CMOS) transistors will be well below ~ 50 nm (Fig. 1), which significantly increases the cost of nanolithography.[1,2] This problem has accelerated the growing interest in new materials and devices, whose properties can be controlled on the nanoscale level, and perhaps even on the molecular level.[3] The use of inherently nanostructured conjugated polymers as components for nanoelectronics is a promising route to future high-density nanochips.[4] Conjugated polymers can be synthesized in a precisely controlled way to form many different nanoscale structures down to the molecular scale. A common feature of conjugated polymers is that their conductivity can be increased by many orders of magnitude upon doping.[5,6] The structural, optical and electrical properties of three-dimensional (3D) conjugated polymers have been studied extensively in the last few decades.[7] Just recently the polymer nanostructures (wires and tubes) made of conjugated polymers such as polyacetylene (PA), polypyrrole (PPy), polyaniline (PANi), polythiophene, and poly(p-phenylenevinylene) have attracted considerable attention (structures shown in Fig. 2). The transport properties of such low-dimensional systems are of great interest because of their potential applications in nanotechnology.[8] The main goal of this work is to provide a brief review of recent advances in the synthesis, electrical-transport characterization, and development of possible device applications of polymer nanofibers and nanotubes.

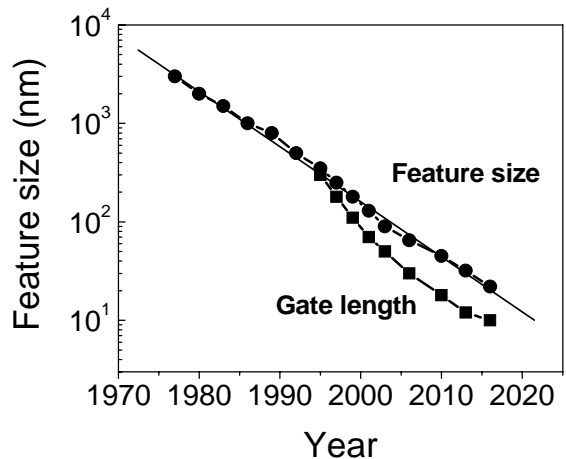

Figure 1. Future technology trends in microelectronics: feature size and gate length.[2]

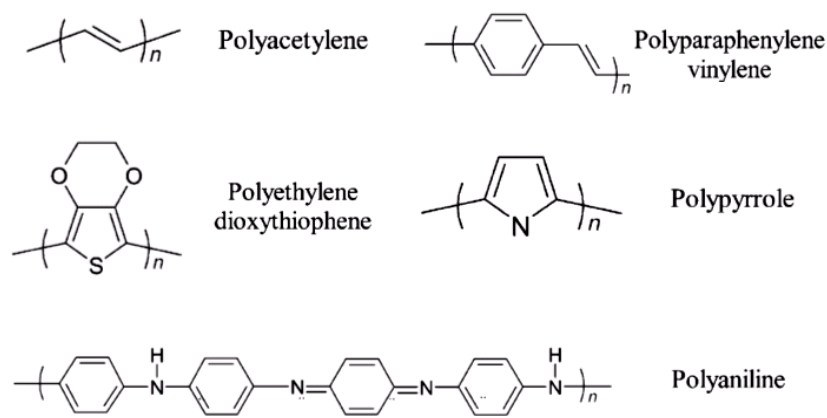

Figure 2. Chemical structure of some of the most important conjugated polymers

## 2. Advances in the Synthesis of Polymer Nanofibers and Nanotubes

There are several general methods to synthesize polymer nanostructures such as fibers and tubes. The most commonly used techniques are template synthesis,[9-11] chiral reaction,[12] self-assembly,[13] interfacial polymerization,[14,15] and electrospinning.[16] The template synthesis method is an effective route to synthesize nanofibrils and nanotubes of various polymers.[9-11] The advantage of the template synthesis method is that the length and diameter of the polymer fibers and tubes can be controlled by the selected porous membrane, which results in more regular nanostructures. The template method has been used to synthesize nanofibers and tubes of PPy, poly(3,4-ethylenedioxythiophene) (PEDOT), PANi and some other polymers. For example, PPy nanotubes have been synthesized using the pores of track-etched polycarbonate membranes as a template; the pore diameter of the template ranged from 50 to 200 nm.[17] PANi and PEDOT nanofibers with a diameters of ~ 75-150 nm were synthesized electrochemically by the same method, and their morphology was studied using electron microscopy.[18] A general feature of the conventional template method is that the membrane should be soluble so that it can be removed after synthesis in order to obtain single fibers or tubes. This restricts practical application of this method and gives rise to a need for other techniques.

It was shown recently that PANi and PPy nanotubes can be synthesized by a self-assembly method without an external template.[19,20] In particular, PANi nanotubes doped with camphor sulfonic acid (CSA) have been obtained which have outer and inner diameters of 175 nm and 120 nm respectively.[14] Interfacial polymerization is another promising method to synthesize PANi nanofibers, where aniline is polymerized to PANi by chemical oxidation at the interface of two immiscible liquids under ambient conditions, without the need for templates or functional dopants.[14, 15] This method leads the formation of PANi fibers with uniform diameters between 30 and 120 nm using hydrochloric and perchloric acid. When CSA is employed, fibers with average diameters of 50 nm and lengths varying from 500 nm to several micrometers are formed. The Brunauer-Emmett-Teller surface area of the nanofibers increases as the average diameter decreases. PANi-AMPSA (AMPSA: 2-acrylamido-2methyl-1-propanesulfonic acid) nanofibers were also synthesized by using this method.[21,22] Interfacial polymerization is shown to be readily scalable to produce bulk quantities of nanofibers. Recently fibers of PANi-CSA blended with polyethylene oxide (PANi-CSA/PEO) were fabricated by an electrostatic spinning (electrospinning) technique.[4,16] It was shown that fiber diameters below 30 nm (near 5 nm) could be obtained with optimized electrospinning process parameters.[23] Scanning conductance microscopy shows that fibers with a diameter below 15 nm are electrically insulating; the small diameter may allow complete dedoping in air or may be smaller than phase-separated grains of PANi and PEO. Nanowires of PEDOT doped with polystyrenesulfonic acid (PEDOT/PSS), a promising polymer for organic devices, with a diameters below 5 nm were prepared recently by a molecular combining method (using atomic force microscope (AFM) lithography).[24]

Among the above-mentioned fibers and tubes, PA nanofibers are of particular interest because of the unique simple chemical structure of this polymer. PA has a bandgap of $E_g \sim 1.5$ eV, and consist of linear chains of CH units with alternating single and double bonds (Fig. 2). Because of its great increase in conductivity upon doping,[5,6] PA fibers are promising for detailed nanoscale transport studies. Networks of PA nanofibers can be prepared by exposing a dilute Ziegler-Natta catalyst to acetylene.[5] Figure 3a shows a scanning electron microscopy (SEM)

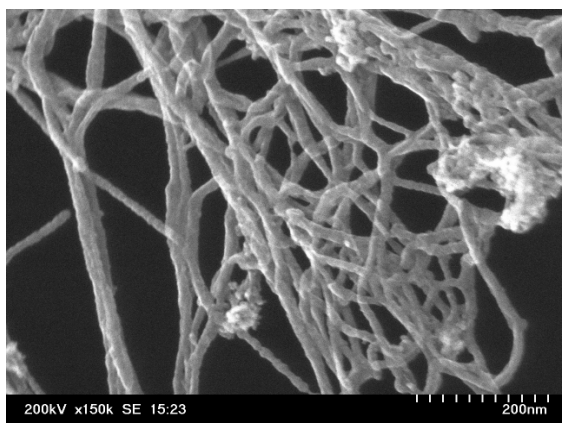
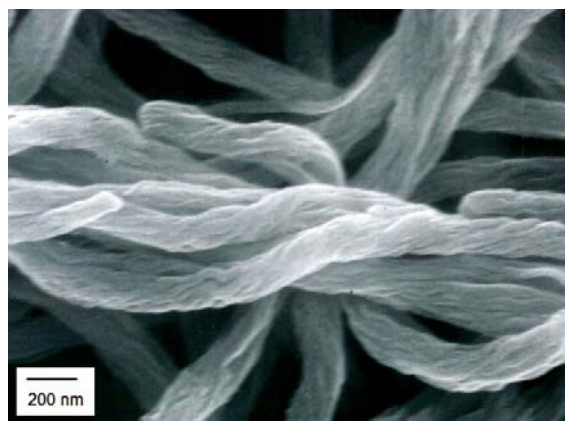

a)                                                                                      b)

Figure 3. SEM images of: a) polyacetylene nanofibers ropes,[25] b) Counterclockwise helical (*R*-helical) polyacetylene nanofibers ropes.[30]

Image of such PA nanofibers ropes.[25] After disintegration of the PA network by ultrasonic treatment, nanofibers with diameters ranging from 100 nm down to 5 nm can be obtained. The properties of such a doped PA network and a single PA fiber were studied recently.[26-28] There is a new, advanced form of PA - helical PA (hel PA), synthesized using a chiral nematic liquid crystal as the solvent for the Ziegler-Natta catalyst.[12] Hel PA can be obtained with either *R*-(counterclockwise) or *S*- (clockwise) type helicity. X-ray diffraction patterns of both *R*- and *S*-type hel PA show a broad reflection, indicating that the films are polycrystalline, corresponding to a spacing of 3.68 A, characteristic of *trans*-PA.[29] A single fiber has a cross-section that is typically 40-65 nm wide and 130-300 nm high. The typical length of hel PA fibers is on the order of 10 μm - much longer than fibers of traditional PA. A SEM image of hel PA ropes and an AFM image of a single *R*-hel PA fiber are shown in Figures. 3b and 5 respectively.[30,31] As can be seen, progress in synthesis enables us to obtain a wide variety of polymer nanofibers and tubes. Knowledge of the transport mechanism in such nanostructures is of importance for future nanoscale device applications.

## 3. Electrical Transport in Polymer Nanofibers and Nanotubes

Charge transport in 3D conducting polymers has been thoroughly studied over the last few decades,[7] but still remains a topic of significant controversy because of the strong influence of disorder on the transport properties of conducting polymers.[32,33] On the one hand heavily doped conjugated polymers demonstrate properties that are characteristic of disordered metals, and the metal-insulator transition (MIT) can be described by a conventional 3D localization-interaction model for transport in disordered metals near the MIT.[32] According to another point of view, the transport properties are dominated by more macroscopic inhomogeneities, and the MIT is better described in terms of percolation between metallic islands.[33] However, even for 3D conducting polymers, there are some transport features (especially at low temperatures) which can not be explained either by localization-interaction or percolation models.[34] Nonetheless, depending on the degree of disorder, one can observe three different transport regimes: metallic, critical, and insulating (as has been summarized previously in the literature).[7,32,35] In the insulating regime of the MIT, conduction occurs by phonon-assisted tunneling between electronic localized states in the band gap. The conductivity shows exponential temperature dependence, $\sigma(T)$, characteristic of variable-range hopping (VRH):[36]

$$\sigma(T) \propto \sigma_0 \exp[-(T_0/T)^p] \qquad (1)$$

where $T_0$ is a VRH parameter, $\sigma_0$ is a prefactor with an algebraic temperature dependence, and p = 1/(d + 1), d being the dimensionality of the system. Thus one obtains p = 0.25, 0.33, 0.5 for three-d, two-, and one-dimensional VRH, respectively. For 3D disordered systems in the critical regime of the MIT, the temperature dependence of the resistivity, $\rho(T)$, follows a universal power-law:

$$\rho(T) \approx (e^2 p_F / \hbar^2)(k_B T / E_F)^{-1/\eta} \approx T^{-\gamma} \qquad (2)$$

where $e$ is the electron charge, $\hbar$ is Planck's constant, $E_F$ is the Fermi energy, $\gamma$ is the power exponent, $\eta = 1/\gamma$, $p_F$ is the Fermi momentum, and $1 < \eta < 3$, i.e. $0.33 < \gamma < 1$.[37] In heterogeneous systems, conduction occurs by electronic tunneling through the non-conducting regions (barriers) separating metallic islands (rather than between localized states). If the metallic islands are large enough, $\sigma(T)$ can be described by a fluctuation-induced tunneling (FIT) model as:[38]

$$\sigma(T) = \sigma_t \exp[T_t/(T + T_S)] \qquad (3)$$

where $T_t$ represents the temperature at which the thermal voltage fluctuations across the tunneling junction become large enough to raise the energy of the electronic states to the top of the barrier, and $T_S$ is the temperature above which thermally activated conduction over the barrier begins to occur. The ratio $T_t/T_S$ determines the tunneling in the absence of fluctuation. For small metallic islands the FIT theory predicts $\sigma(T) \sim -T^{-0.5}$ similar to one-dimensional (1D) VRH. A thorough analysis of the applicability of different transport models to 3D conducting polymer films has been presented previously.[35]

Charge transport in low-dimensional structures such as polymer nanofibers and nanotubes is less studied than for in 3D conducting polymer films. Experimental data is available on the electrical properties of PA,[26-28] PPy,[17] PEDOT,[18] PANI,[16,20,23] and some other polymer fibers and tubes. The results for different fibers are not uniform and there is no generally accepted model for a dominant transport mechanism in such systems. In particular, the current-voltage (I-V) characteristics of PPy nanotube (~ 120 nm in diameter) deposited on Au were studied using a metal-coated tapping-mode AFM tip.[17] The observed linear I–V characteristics lead to an estimation of the room temperature electrical conductivity ($\sigma_{300K}$) of PPy nanotube as $\sigma_{300K} \sim 1$ S/cm, which is consistent with $\sigma_{300K}$ values obtained for bulk PPy films.[32] The temperature dependence of the conductance, G(T), of PEDOT nanofibers (~ 75 -150 nm in diameter) synthesized by template method was investigated from 70-300 K and the results were compared with those of PEDOT thin films.[18] It was found that for PEDOT fibers ($\sigma_{300K} \sim 37$ S/cm) the G(T) is greater than for PEDOT films, and this variation increases for smaller fiber diameters. Figure 4 shows that, even though G(T) for PEDOT nanofibers can be approximately fitted by Eq. (1) with p ~ 0.25, in fact, it is quite close to the critical regime of the MIT, and follows a power law similar to PEDOT films.[39] These results suggest that there is a confining effect on the PEDOT structure occurs during electrochemical synthesis in the template. As for the PEDOT/PSS nanowires synthesized by AFM lithography, the $\sigma_{300K}$ ~ 0.07 - 0.15 S/cm[24] , which is also on the same order of magnitude as the PEDOT/PSS films reported earlier.[40] The $\sigma(T)$ of a single PANi-CSA nanotube (175 nm in outer diameter) synthesized by a template-free self-assembled method was studied by a standard four-terminal technique.[20,41] The conductivity of the single PANi nanotube is found to be rather high, $\sigma_{300K}$ ~ 31.4 S/cm. The intrinsic resistance of an individual nanotube (30 k$\Omega$) was found to be much smaller than the contact resistance of crossed nanotubes (500 k$\Omega$) and that of PANi nanotubes obtained by the template method.[20] As can be seen from Fig. 4, the G(T)/G(300K) dependence of the PANi-CSA nanotube follows 3D VRH model: $\ln\sigma(T)$ ~ $-T^{-0.25}$.[36] This result means that the presumed low-dimensional PANi-CSA tubes behave as 3D systems, which implies a high degree of disorder in these samples. In the case of a PANi/PEO single fiber fabricated by an electrospinning technique, I–V characteristics show that the thin fibers conduct more poorly than thick ones.[23] Thus, PANi/PEO fibers with diameters of 70 nm and 20 nm have conductivities of ~ $10^{-2}$ S/cm and ~ $10^{-3}$ S/cm, respectively. The I–V characteristics of the PANi/PEO fibers are rectifying, consistent with the formation of Schottky barriers at the nanofiber-metal contacts. There has been no information on the $\sigma(T)$ behavior of PANi or PANi/PEO single fibers obtained by electrospinning techniques or interfacial polymerization untill now. Charge transport in doped PA nanofibers, however, has been studied more thoroughly.[26-28] The $\sigma_{300\ K}$ of iodine doped PA fiber networks and PA single fibers were found to be ~ 420 S/cm and ~ 0.1 S/cm, respectively. The temperature dependence of the I-V characteristics of a doped single PA fiber was studied from 300 – 30 K. The results demonstrate the absence of any I-V temperature dependence below 10 - 30 K with negligible low (~ 0.1 %),[26] or almost zero[28] magnetoresistance (MR) in magnetic fields up to 7 T. The authors have discussed the obtained data in the framework of either a Zener-type tunneling model or a novel conduction mechanism based on the tunneling of a segment of the conjugated bond system in the presence of an electric field by analogy to the soliton-pair creation mechanism in charge-density-wave (CDW) materials.[42,43] However, in our opinion, both the Zener-type or soliton tunneling models suggested for doped PA fibers are still incomplete and require additional G(T), MR and electron-spin resonance studies to be proved. One of the most promising ways to increase nanofiber conductivity is the modification of the polymer fiber itself during synthesis, followed by subsequent doping. As a result of such treatment, the polymer fiber can be twisted in different directions and thus its structural and electrical properties can be modified. This approach has been used in hel PA fibers synthesized by K. Akagi et al.[12] as described in Section 2. The unusually high length of these fibers (up to 10 µm) and their relatively high conductivity ($\sigma_{300K}$ ~ 1 S/cm) enables a detailed study of charge transport in hel PA films[44] and fibers[31,45] down to low temperatures (as discussed below). The geometry, conductance, and conductivity for some of these above-mentioned polymer fibers and tubes are summarized in Table 1.

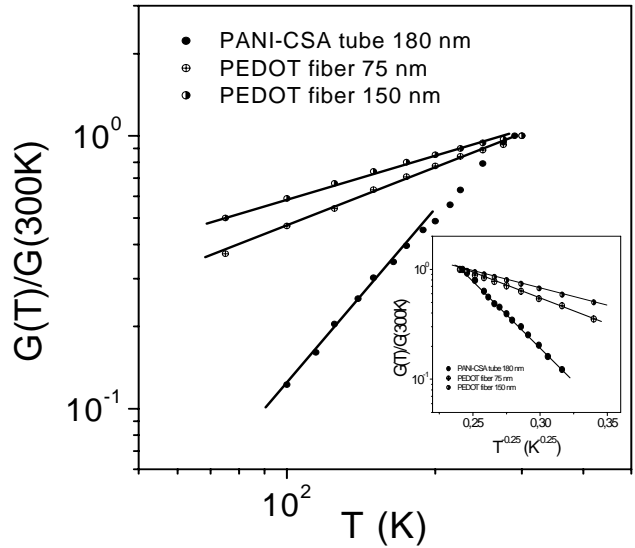

Figure 4. G(T)/G(300 K) vs T for PANi-CSA tube and PEDOT fibers (our treatment of data from [18, 20]). Inset shows lgG(T)/G(300 K) vs $T^{-0.25}$ for the same samples.

**Table 1. Parameters for polymer fibers and tubes**

| # | Sample | Diameter or cross-section, nm | $G_{300 K}$ S | $\sigma_{300 K}$ S/cm | $\alpha$ | $\beta$ (at min T) | Reference |
|---|--------|-------------------------------|----------------|------------------------|----------|---------------------|-----------|
| 1 | R-hel PA fiber | 65×290 | $8.4\ 10^{-7}$ | 1.13 | 2.2 | - | [31] |
|   |        |        |        |        | 2.8 | 2.5 (30 K) | [31] |
| 2 | R-hel PA fiber | 60×134 | $1.1\ 10^{-7}$ | 0.85 | 5.5 | 4.8 (50 K) | [31] |
| 3 | R-hel PA fiber | 47×312 | $2.1\ 10^{-9}$ | 0.0036 | 7.2 | 5.7 (95 K) | [31] |
| 4 | Single PA fiber | 20 | $7.3\ 10^{-9}$ | 0.01 | 5.6 | 2.0 (94 K) | [26] |
| 5 | PPy tube | 15 | $1.7\ 10^{-8}$ | 0.83 | 5.0 | 2.1 (56 K) | [46] |
| 6 | PPy tube | 50 | $2.8\ 10^{-8}$ | 0.83 | 4.1 | 2.8 (50 K) | [46] |
| 7 | R-hel PA 4 fibers | - | $2\ 10^{-8}$ |  | 3.7 | 2.3 (90 K) | [31] |
| 8 | PANI tube | 180(140) |  | 31.4 |  |  | [41] |
| 9 | PANI/PEO fiber | 20 |  | 0.001 |  |  | [23] |
| 10 | PANI/PEO fiber | 70 |  | 0.01 |  |  | [23] |
| 11 | PEDOT fiber | 75 |  | 37 |  |  | [18] |
| 12 | PEDOT/PSS fiber | ~ 5 |  | 0.07-0.15 |  |  | [24] |

One can summarize by saying that despite the rapid growth in the amount of experimental data on the electronic properties of single polymer fibers and tubes, results discussing the $\sigma(T)$ behavior are limited. In most cases, the $\sigma(T)$ has an activated nature and is described by ether 3D VRH or FIT models, likely because of highly inhomogeneous or amorphous structure of the polymer nanofibers under consideration. It is quite within reason to suggest that in polymer fibers and tubes with a higher degree of crystallinity, their inherent 1D nature would be clearly manifested in the transport properties.

## 4. 1D Nature of Conjugated Polymers and Theories for Tunneling in 1D Conductors.

### 4.1 1D Transport in Inorganic Nanowires - Luttinger Liquid Model

In this section, the transport features of polymer nanofibers and tubes, characteristic of 1D systems will, be discussed briefly. It is known that electron-electron interactions (EEIs) strongly affect transport in 1D systems by leading to phases different from conventional Fermi liquid. Specifically, repulsive short-range EEI result in Luttinger liquids (LLs),[47] while long-range Coulomb interactions (LRCIs) lead to Wigner crystal (WC).[48] A characteristic feature of 1D systems is the power-law behavior of the tunneling density of states near the Fermi level, which manifests itself in the power-law temperature dependencies of the conductance, G(T), and the I-V characteristics, I(V). For a clean LL (a Luttinger liquid state related to a 1D conductor without impurities and without interactions with other LLs), a power-law variation of G(T) is predicted at small biases (eV << $k_BT$), G(T) ∝ $T^\alpha$ and a power-law variation of I(V) - at large biases (eV >> $k_BT$), I(V) ∝ $V^\beta$, where the exponents of the power laws depend on the number of 1D channels. The LL state survives for a few 1D chains coupled by Coulomb interactions, and can be stabilized in the presence of impurities for more than two coupled 1D chains.[49,50] According to the LL theory, I-V curves taken at different temperatures should be fitted by the general equation:[51]

$$I = I_0\, T^{\alpha+1} \sinh(eV/k_BT)\, \big|\, \Gamma(1 + \beta/2 + i\, eV/\pi k_BT)\,\big|^2 \qquad (4)$$

where Γ is the Gamma function, V is the voltage bias, and $I_0$ is a constant. The parameters α and β correspond to the exponents estimated from the G(T) and I(V) plots. Equation 4 implies that the I-V curves should collapse into a single curve, if $I/T^{\alpha+1}$ is plotted as a function of $eV/k_BT$. These power-law variations have recently been found to be valid for inorganic 1D systems including single wall carbon nanotubes (SWNT), multi-wall carbon nanotubes (MWNT) (α, β ~ 0.36),[52,53] and doped semiconductor nanowires of InSb (α ~ 2-7, β ~ 2 − 6)[54] and NbSe$_3$ (α ~ 1 - 3, β ~ 1.7 - 2.7),[55] and for fractional quantum Hall edge states in GaAs (α, β ~ 1.4 - 2.7).[56] The LL, WC or Environmental Coulomb blockade (ECB)[53] models are currently debated in order to explain the above-mentioned power-law behavior in 1D systems. It is worth noting that nanofibers made of conjugated polymers are essentially another example of initially 1D systems.[5,6] Recent results on transport in conventional and helical single PA fibers and PPy nanotubes demonstrate that for all these polymer nanofibers and tubes, both G(T) and I(V) show power-law behavior characteristic of 1D systems such as LL.[31,45] These results are in greater detail in the subsequent sections.

## 4.2 Recent Transport Experiments on Polymer Nanofibers and Nanotubes

Charge transport in iodine doped *R*-hel PA fibers synthesized by K. Akagi[12] has been studied in the temperature range from 30 to 300 K.[31] The details of a single *R*-hel PA fiber dispersion are described elsewhere.[45] Figure 5 shows the AFM image of a *R*-hel PA fiber on top of Pt electrodes. The typical G(T) for *R*-hel PA nanofibers, as well as the results of our analysis of early published data for iodine doped single PA nanofibers (~ 20 nm in diameter) and PPy nanotubes (~ 15 nm in diameter),[26,46] are presented in Figure 6 and Table 1. As can be seen from Figure 6, the conductance for all types of polymer fibers show a power-law behavior, G(T) ∝ T$^\alpha$, starting from room temperature, and down to ~ 30 K for the most conductive *R*-hel PA nanofiber. The power exponent increases from ~ 2.2 up to ~ 7.2 as the fiber diameter or its cross-section (the amount of polymer chains) becomes smaller. The I-V characteristics for the *R*-hel PA fiber at low temperature also follow the power law I(V) ∝ V$^\beta$ (Fig. 7). The same power-law variations for both G(T) and I(V) are found for all other PA fibers and PPy tubes and even for four crossed *R*-hel PA fibers, with a variety of power exponents $\alpha \sim 2.2 - 7.2$ and $\beta \sim 2 - 5.7$. The variation of $\alpha$ and $\beta$ from one fiber to another may result from the scatter in nanofiber diameters or from variations in the doping level. It has been previously suggested that the power-law variations in G(T) and I(V) are a general feature of charge transport in polycrystalline polymer nanofibers and nanotubes.[31] Note that similar power-law variations are found for inorganic 1D systems, including SWNT and MWNT, InSb and NbSe$_3$ nanowires, and quantum Hall edge states in GaAs.[52-56] It is evident that polymer nanofibers and nanotubes differ from 1D carbon nanotubes and semiconductor nanowires. Each $R$-hel PA fiber consists of a number of oriented 1D chains, which are to some extent coupled and disordered as a result of doping. Therefore, polymer fibers are in fact quasi-1D, and the confinement effects are not expected to be significant. However, the remarkable similarities to inorganic 1D nanowires observed in G(T) and I(V) behavior of polymer fibers leads us to examine various theoretical models for tunneling in 1D systems.

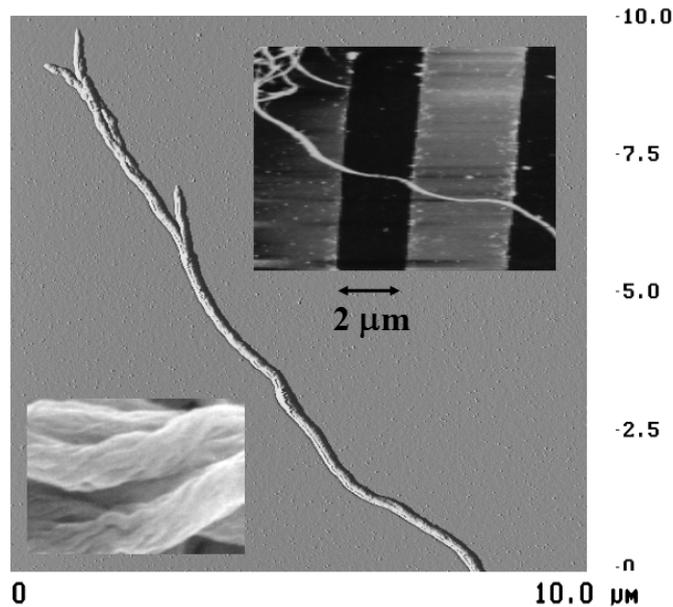

Figure 5. AFM image of *R*-hel PA fiber. Insets show the *R*-hel PA fiber on top of Pt electrodes (top) and SEM image of *R*-hel PA microfibers (bottom).

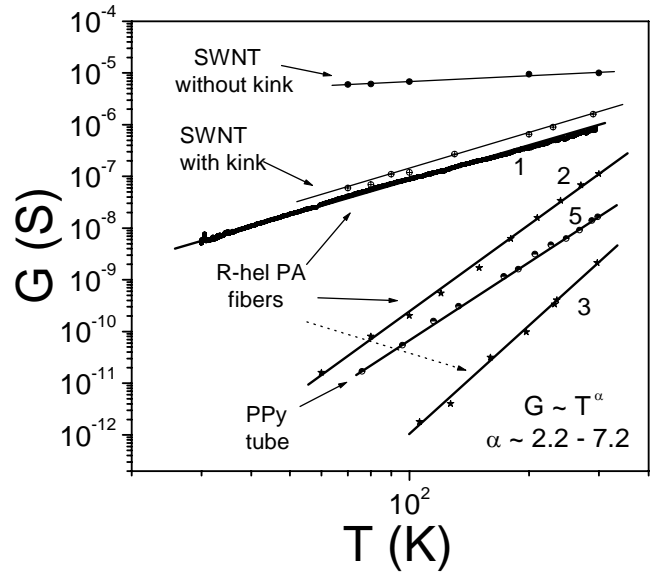

Figure 6. Conductance vs T for different polymer fibers and tubes in comparison with G(T) for a SWNT with a kink from [66].

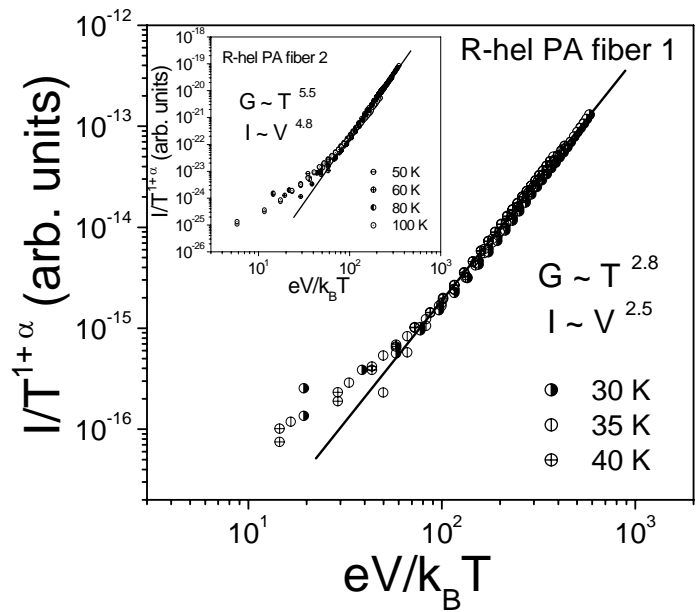

Figure 7. I/T $^{\alpha+1}$ vs eV/$k_B$T for $R$-hel PA fibers 1 and 2 (inset) from Table 1 at different temperatures, where $\alpha$ is the exponent in G(T) $\propto$ T$^{\alpha}$.

## 4.3 Applicability of Different 1D Tunneling Models to Polymer Nanofibers and Nanotubes

The polymer fibers and tubes under consideration are initially 1D conductors; thus, one can expect that either short-range EEIs or LRCIs will affect transport in such quasi-1D systems. However, the applicability of some traditional models should also be considered. First, one can analyze the G(T) curves in the framework of 1D VRH[36,57,58] and FIT[38] models. The VRH model implies that $\sigma$ (T) should be along the lines of Equation 1, i.e. G(T) $\propto$ G$_0$ exp[-(T$_0$/T)$^p$]. The standard derivation of VRH transport for d-dimensional hopping predicts the power exponent p in Equation 1 to have the form: p = ($\gamma$ + 1)/($\gamma$ + d + 1).[36,57] In this expression $\gamma$ is the power exponent in a power-law density of states g($\varepsilon$) $\propto$ $\varepsilon^\gamma$ and d is the dimensionality of the system. As can be seen, the exponents p of VRH conductivity for 1D systems are strongly dependent on the shape of the density of states g($\varepsilon$) near the Fermi level E$_F$, which arises due to 3D Coulomb interactions. Thus, for d = 1, one obtains p = 0.5, 0.67, and 0.75 by setting $\gamma$ to 0, 1, and 2 respectively. However, even on fitting with p ~ 0.25, G(T) is too strong in polymer fibers to be explained by the 3D VRH model. Note that one may obtain power law behavior for G(T) with p ~ 0.25 (3D VRH), assuming that the pre-exponent G$_0$(T) is temperature dependent.[57] This approach works over a rather narrow interval for G(T) and is not valid for our polymer fibers where the power-law behavior is observed for several decades of G(T). The latter pure FIT model reveals unreasonable fitting parameters in fitting the power law behavior of G(T).[46] A tunneling transport mechanism based on the FIT model has been recently considered for PA fibers,[42,43] however, this approach does not take into account the 1D nature of the polymer fibers. It is known that for a 3D disordered systems, in the critical regime of the MIT, the $\rho$(T) follows an universal power-law, as described by Equation 2, with $0.33 < \gamma < 1$.[37] This model with $\gamma < 1$ can explain charge transport in non-oriented 3D *R*-hel PA films at T > 30 K.[44] For quasi-1D polymer nanofibers, all the exponents are above unity, and therefore this model can be ruled out. Power law behavior, $I(V) \propto V^{\beta}$, with $\beta \approx 2$ is seen for a space-charge limited current (SCLC) transport mechanism in semiconductors.[59] The values $\beta > 2$ observed for *R*-hel PA fibers and PPy tubes thus render the SCLC model inapplicable here. The analysis detailed above has motivated us to consider the applicability of other theories for tunneling in 1D conductors to polymer nanofibers.

As mentioned above, the power-law variations of G(T) and I(V) in inorganic 1D systems, such as SWNTs, MWNTs, and doped semiconductor nanowires are discussed in terms of LL, WC or ECB theories.[52-56] According to the LL or ECB models, all the I-V curves taken at different temperatures should be fitted by the general Equation 4, i.e. they should collapse into a single plot $I/T^{\alpha+1}$ versus $eV/k_BT$. As can be seen from Figure 7, all the scaled I-V curves for *R*-hel PA fibers collapse into a single curve at low temperatures by plotting $I/T^{\alpha+1}$ versus $eV/k_BT$, where $\alpha$ is the power-law exponent estimated from the G(T) curve for the same sample. A similar scaled behavior is found for all other *R*-hel PA fibers, as well as for a single PA fiber and PPy tubes.[31] The exponents $\alpha$ and $\beta$ (at the lowest possible temperatures) are listed in Table 1. One can again conclude that the power-law variations in G(T) and I(V), as well as the scaled I-V behavior, are characteristic for quasi-1D systems such as polymer fibers and tubes. To a certain degree, this behavior correlates with LL theory predictions for transport in 1D systems. This theory implies that both tunneling along LLs through impurity barriers and tunneling between chains with LLs provide for conduction of a set of coupled LLs. The system is characterized by measuring the LL interaction parameter, g, from the tunneling density of states.[52] LL theory predicts the conduction exponents α as being (1/g − 1)/2 for impurity barriers, and (g + 1/g − 2)/4 for interchain tunneling.[60] The relative contribution of these parallel conduction channels depends on the strength of impurities, temperature and electric field. The latter equation for the interchain tunneling allows g to be estimated as ~ 0.08 for the most conductive *R*-hel PA sample. This is much lower than g ~ 0.2 reported for tunneling into a SWNT from metal electrodes.[51,52] The low g value indicates that strong repulsive EEIs, which are characterized by g << 1, affect transport in polymer fibers, whereas for non-interacting electrons g = 1. It is noteworthy that the LL model for a single chain requires a correlation between the exponents so that β = α + 1. However, as can be seen from Table 1, for all the polymer fibers under consideration β ≠ α + 1, and, moreover, β is always less than α, which can not be explained by the LL or ECB theory.[53] The same disagreement with a pure LL model has been found for inorganic 1D nanowires[54,55] and can be associated with the fact that a LL in the presence of disorder may not exhibit features typical for a clean case.[61] The interchain interactions may cause a renormalization of the exponents at high temperatures, whereas interchain hopping destroys the LL state at low temperatures. Since interchain hopping is one of the most dominant transport mechanism in doped conjugated polymers at low temperatures,[7] one can expect the manifestation of the LL state in doped polymer fibers only at relatively high temperatures. This correlates with the results obtained for *R*-hel PA nanofibers at T > 30 K.[31]

On the other hand, repulsive LRCIs are believed to lead to the formation of a WC, which is pinned by an impurity.[48] According to theory, WCs occur in solids with a low electronic density, i.e. with a large parameter $r_S = a/2a_B$, where *a* is the average distance between electrons and $a_B$ is the effective Bohr radius. This corresponds to a negligible quantum fluctuation and can be rewritten as $r_S = E_C /E_F$, where the Coulomb energy $E_C$ is larger than the kinetic energy $E_F$ of the electrons. The WC transition has been shown to take place at $r_S$ ~ 36.[62] Such a pinned WC can adjust its phase in the presence of disorder to optimize the pinning energy gain, similar to classical CDW systems. Polymer nanofibers and tubes are quasi-1D conductors with a low electronic density where WC may occur. As the molecular structure of PA consists of alternating single and double bounds in the CH chain, this results in a Peierls distortion, which creates potential wells where the electrons can get trapped, and hence form WCs (Fig. 8). At a modest doping level of the PA chain, the distance between impurities (iodine) is of the order of 10 nm, so $r_S$ is large enough to obtain a WC in the PA nanofiber. When the localization length is larger than the distance between impurities, the tunneling density of states follows a power law[48,63] with high values for the power exponents, ~ 3 - 6,[64] similar to those found for G(T) in polymer nanofibers. However, there are strong arguments against the applicability of a WC model to polymer fibers. First, the impurities in the doped conjugated polymers are located outside of the polymer chains, and therefore they only supply charge carriers without really pinning the chain. This results in the polymer chain either not being pinned or being pinned weakly by conjugated defects. Next, for a classical 1D WC pinned by impurities, one should expect a distinct VRH regime at low temperatures with an exponential G(T).[65] The absence of effective pinning prevents the observation of VRH in polymer fibers, at least at T > 30 K, which argues against the applicability of the WC theory to explain the polymer nanofiber results. As compared with LLs, WCs are even more strongly affected by disorder, and can not survive in a doped polymer system.

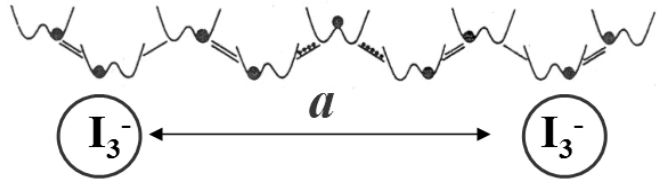

Figure 8. Polyacetylene chain with dopant (iodine) atoms.

Therefore, the analysis above demonstrates that the pure VRH, LL, ECB or WC models cannot precisely describe the power-law variations observed for G(T) and I(V) in polymer nanofibers, despite the fact that some similarities are observed with the LL and WC models. It was supposed that the real transport mechanism in quasi-1D polymer fibers obeys ether a superposition of the above-mentioned models or follows a single LL-like model valid for different parts of the metallic polymer fiber separated by intramolecular junctions.[31] By analogy with LL transport in bent metallic carbon nanotubes,[66] it has been suggested that the conductance across an intramolecular junction is much more temperature dependent than across the two (or more) straight segments, but still obeys the power law, $G(T) \propto T^\alpha$. In the case of SWNTs, for end-to-end tunneling between two LLs separated by an intramolecular junction (kink), the power exponent $\alpha_{\text{end-end}} = (1/g - 1)/2 \sim 1.8$, if the LL interaction parameter $g \sim 0.22$. Figures 6, 9 demonstrate that for a SWNT with a single kink, the exponent $\alpha$ is 2.2,[66] which is close to the calculated value and surprisingly close to the $\alpha$ value for the most conductive *R*-hel PA fiber 1. It is evident from AFM studies that each *R*-hel PA fiber contains kinks when the distance between the electrodes is 2 μm (Figs. 5, 9). Each kink in the polymer fiber acts as a tunneling junction between the ends of two LLs. Therefore, each fiber can be approximated as a system of LL parts, connected in series by intramolecular junctions, with a power-law behavior for the density of states near the Fermi level in each part. This effect results in stronger G(T) and I(V) power-law variations with respect to a clean LL, and may cause the above-mentioned contradictions with a pure LL model. The similar strong power-law behavior of G(T) and I(V) observed for the four crossed *R*-hel PA fibers is reminiscent of LL behavior in crossed metallic SWNT[67] and provides further support for this model. Recent structural studies of hel PA fibers by SEM reveals that each fiber is composed of helical micro-fibers that are small bundles of polymer chains (Fig. 3b, Fig. 5 inset).[30] This is different from conventional, non-helical PA (Fig. 3a), and may affect the transport in a more complicated way by leading to a higher power exponent for G(T) and I(V). One can expect that in such microfiber systems, long-range interactions between electrons in each microfiber may be screened through Coulomb interaction of these electrons with electrons of neighboring microfibers. This leads to a short-range intrafiber EEI, which is the basic assumption of the LL theory. In all other respects, the above described model of LLs separated by intramolecular junctions should be valid for the microfibers case. Therefore, one can summarize that the power-law behavior observed in *R*-hel PA fibers for both G(T) and I(V) is characteristic of such 1D systems comprising several LL parts connected in series. At the same time, there is a discrepancy between our results for polymer fibers and theories for tunneling in truly 1D systems. Further effort is necessary to clarify the origin of the power-law behavior for G(T) and I(V) observed in polymeric and inorganic 1D systems.

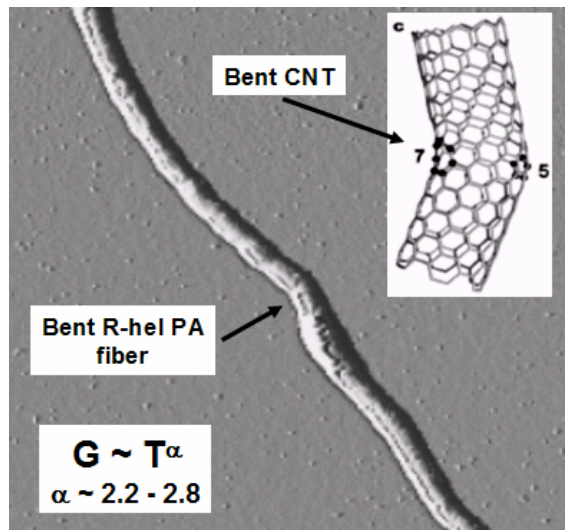

Figure 9. *R*-hel PA fiber and carbon nanotube with kinks.

## 5. Device Applications of Polymer Nanofibers and Nanotubes

The application of polymer nanofibers and nanotubes as elements of nanoelectronic devices is a worthy challenge for nanotechnology. For example, thin films made from PANi nanofibers synthesized by interfacial polymerization show superior performance as gas sensors in terms of both sensitivity and time response to vapors of acid (HCl) and base (NH$_3$).[14,15] Conducting PEDOT nanowires (~ 200 nm in diameter) synthesized by an electrochemical polymerization method using Al$_2$O$_3$ nanoporous templates have the potential for use as nanotips in field-emission displays.[68] It has been demonstrated that the electrical properties of PA fibers (~ 20 nm in diameter) can be efficiently studied using a field-effect transistor (FET) geometry by changing their conductivity through the application of a gate voltage.[27] Figure 10a shows the schematic diagram of such a FET based on a single PA fiber. Similar FET techniques have also been used to study electronic transport in doped PPy[17] and PANi/PEO nanofibers.[23,69] It was found that the I-V characteristics of polymer single fiber FETs are strongly non-linear (Fig. 10b). This is different from semiconductor[70] and organic FETs,[71,72] which show saturation behavior. At the same time, this behavior is similar to the behavior seen in SWNT,[73]

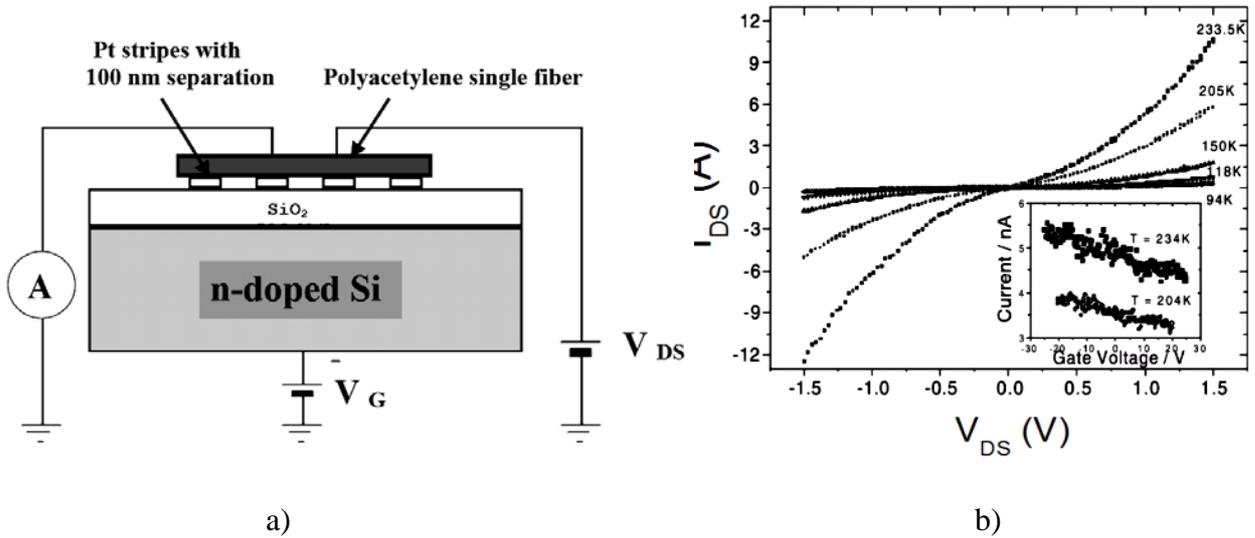

a)                                                                b)

Figure 10. a) Schematic diagram of a PA single fiber FET. b) Typical I-V characteristics of a PA nanofiber FET (D ~ 20 nm) at different temperatures. Inset shows the transfer curves for the same fiber at $V_{DS}$ = 1.0 V.[27] Copyright of the Elsevier 2001. Reproduced from [27] with permission.

poly(2-methoxy-5-(2'ethylhexoxy)-1,4phenylenevinylene) (MEH-PPV) thin strip FETs,[26] DNA-templated SWNT FETs,[74] and some other low-dimensional structures.[3] To evaluate the field effect mobility ($\mu_{FET}$) of polymer nanofiber one can consider the relatively low voltage region of the I-V characteristics in terms of the linear regime in conventional metal oxide semiconductor field-effect transistors (MOSFETs). Then from the transconductance, $g_m$, the $\mu_{FET}$ can be estimated as:[70]

$$g_m = \partial\, I_{DS}\, /\, \partial\, V_g\, |_{Vds\,=\,const} = -\,(Z\, /\, L)\, \mu_{FET}\, C_i\, V_{DS} \qquad (5)$$

where Z is the channel width, L is the channel length and $C_i$ is the capacitance per unit area. For the PA single fiber FET the $\mu_{FET}$ at 233 K was found to be $\mu_{FET}$ ~ 1.5 $10^{-3}$ cm$^2$/Vs (L ~ 100 nm, D

~ 4.4 nm, $C_i$ ~ 12 nF/cm$^2$).[27] The observed non-linear I-V characteristics of PA single nanofiber FETs can be explained by either SCLC[59] or Schottky-barrier[70] mechanisms. The presence of traps and non-optimal contact barriers, as well as a rather large channel length, results in rather low $\mu_{FET}$ values in PA nanofiber FET structures. In order to increase the electrical stability and FET mobility performance, significant attention has recently been focused on PANi nanofibers. In particular, conducting polymer PANi nanofibers made by electrodeposition and self-assembly methods have been under intensive discussion.[16,20,23] It is worth noting that electrospinning is a well-established approach to polymer fiber fabrication with the possibility for large scale production of long polymer fibers for incorporation into smart textiles and wearable electronics.[75] It has been shown that electrospinning is a promising method to obtain PANi-based nanocables.[16,23,76] In addition, FET behavior has been reported recently for PANi-CSA/PEO nanofibers made by electrospinning.[69] Saturation channel currents were observed at surprisingly low source–drain voltages of -0.4 - 0.6 V, while the 1D charge density (at zero gate bias) was calculated to be about one hole per 50 two-ring repeat units of PANi, consistent with the rather high channel conductivity (~$10^{-3}$ S/cm). It has been found that for PANi-CSA/PEO nanofibers, the hole $\mu_{FET}$ in the depletion regime is ~ $1.4 \times 10^{-4}$ cm$^2$/Vs - lower than for FET structures based on PA fibers. Reducing or eliminating the PEO content in the fiber is expected to enhance device parameters. The relatively low $\mu_{FET}$ values observed for FETs based on PA, PPy, and PANi/PEO polymer fibers have motivated us to look for polymers of better quality, as well as to strive for improved geometries in nanofiber FETs. To make polymer fiber FETs more efficient, one should decrease the amount of traps as well as the concentration of conjugation defects during polymer synthesis. A higher degree of crystallinity of the polymer fibers should also lead to better device performance. Another challenge for polymer nanofiber FETs is the search for a cost-effective way to position the fibers and tubes (from solution) in a given direction in a high-density electronic circuit. In parallel, one should also try to decrease the channel length down to several nm. The use of polymer fibers with a diameter on the order ~ 10 nm in ultimate double-gate FET structures (Fig. 11a) in combination with high-dielectric constant (k) gate dielectrics, should lead to a significant improvement in device characteristics. The proposed tubular structure shown in

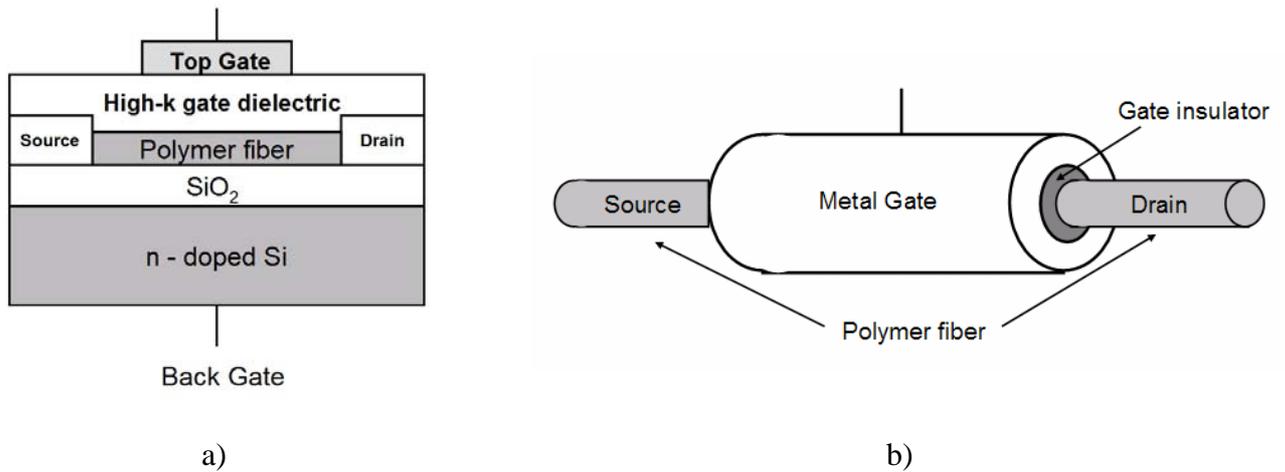

a)                                                    b)

Figure 11. a) Schematic of a double-gate FET based on a polymer fiber. b) An unrealized, ideal nanofiber FET.

Figure 11b is almost ideal in terms of electrostatic control of charge carriers in the channel if it can be coupled with a well-controlled high-k gate insulator and a metal gate with an appropriate workfunction. Such a presumed FET structure can be realized by either a carbon nanotube or by a polymer nanofiber.[2] Understanding the transport mechanism in such an ideal nanowire FET structure would be a worthwhile endeavor. One can expect that by using a conjugated polymer and organic molecular wires it might be possible to combine the insulating and semiconducting properties of such low-dimensional organic materials to achieve enhanced nanodevice characteristics.

## 6. Conclusions

There has been significant progress in the synthesis and electrical characterization of nanofibers and tubes made of conjugated polymers during the last decade. Understanding of electrical transport in polymer nanofibers and tubes has improved dramatically, but still remains the subject of intensive discussion. Recent advances in electrical properties and device performance characteristics have resulted from the introduction of novel, more stable polymeric materials and methods promising for applications in nanotechnology. The applicability of various theoretical models has been considered in order to understand the recent experimental results on transport in conducting polymer nanofibers and tubes. The power-law behavior observed recently in polymer fibers and tubes, for both G(T) and I(V), is characteristic of such 1D systems composed of several LLs connected in series. This result indicates that the inherent 1D nature of conjugated polymers clearly manifests itself in the transport properties of conducting polymer nanofibers and tubes. However, some discrepancy between our results and theoretical models for tunneling in truly 1D systems are found, especially at low temperatures. This underlines the need for further experimental and theoretical efforts to clarify the origin of low temperature transport in such quasi-1D systems as conducting polymer nanofibers and tubes. Some promising nanoelectronic applications of polymer nanofibers and tubes as double-gate and tubular structure FETs have been suggested in order to achieve better device performance in the near future.

## Acknowledgements

The author is grateful to Y. W. Park, H. J. Lee, and V. I. Kozub for collaboration and to K. Akagi for sending the SEM picture. This work was supported by the Nano Systems Institute - National Core Research Center (NSI-NCRC) program of KOSEF, Korea. Support from the Brain Pool Program of Korean Federation of Science and Technology Societies for A. N. A. is gratefully acknowledged.